\documentstyle[12pt]{article}

\def\bseq{\begin{subequation}} 
\def\eseq{\end{subequation}}
\def\bsea{\begin{subeqnarray}} 
\def\esea{\end{subeqnarray}}

\def\beq{\begin{equation}}
\def\eeq{\end{equation}}
\def\eea{\end{eqnarray}}
\def\bq{\begin{quote}}
\def\eq{\end{quote}}

\newcommand{\EQ}{\begin{equation}}
\newcommand{\EN}{\end{equation}}
\newcommand{\bea}[1]{\begin{eqnarray}\label{#1}}
\newcommand{\ena}{\end{eqnarray}}

\renewcommand{\a}{\alpha}

\renewcommand{\d}{\delta}
\newcommand{\th}{\theta}

\newcommand{\G}{\Gamma}

\newcommand{\e}{\epsilon}

\newcommand{\m}{\mu}

\newcommand{\p}{\pi}

\newcommand{\s}{\sigma}

\newcommand{\f}{{\phi}}
\newcommand{\fb}{\bar {\phi}}
\newcommand{\phibar}{\bar{\phi}}
\newcommand{\Dbar}{\bar{D}}
\newcommand{\FF}{{\cal F}}

\newcommand{\Phibar}{\bar{\Phi}}

\newcommand{\shalf}{\hbox{$\frac12$}} 

\def\bop#1{\setbox0=\hbox{$#1M$}\mkern1.5mu
 \vbox{\hrule height0pt depth.04\ht0
 \hbox{\vrule width.04\ht0 height.9\ht0 \kern.9\ht0
 \vrule width.04\ht0}\hrule height.04\ht0}\mkern1.5mu}

\def\eed{{\hbox{C\kern-.3em I}}}

\begin{document}
\begin{titlepage}
\begin{flushright}
IFUM-568-FT\\
ITP-SB-97-28\\
BRX-TH-416\\
USITP-97-09\\
hep-th/9706013\\
\end{flushright}

\begin{center}
{\large\bf  THE N=2 SUPER
YANG-MILLS  LOW-ENERGY EFFECTIVE ACTION AT TWO LOOPS}\\
\vspace{0.5cm}

\normalsize
A. De Giovanni \footnote{email:degiovanni@pv.infn.it}\\{\it Dipartimento di
Fisica, Universit\`a di Pavia and INFN,}\\
{\it Sezione di Pavia, Via Bassi 6, I-27100 Pavia, Italy.}\\
\bigskip
M.T. Grisaru\footnote{email:grisaru@binah.cc.brandeis.edu}\\{\it
Physics Department,  Brandeis University}\\
{\it Waltham, MA 02254, USA.}\\
\bigskip
M.\ Ro\v cek\footnote{email:rocek@insti.physics.sunysb.edu}\\
{\it Institute for Theoretical Physics,  State University of New York}\\
{\it Stony Brook, NY 11794., USA.\\}
\bigskip
 R.\ von Unge\footnote{email:unge@vanosf.physto.se}\\
{\it Department of Physics, Stockholm University\\}
{\it Box 6730, S-11385 Stockholm, Sweden.\\}
\bigskip
 D. Zanon \footnote{email:zanon@mi.infn.it} \\
{\it Dipartimento di Fisica, Universit\`a di Milano and
INFN, }\\
{\it Sezione di Milano, Via Celoria  16,  I-20133 Milano, Italy.}

\end{center}
\vspace{0.2cm}
\begin{abstract}
We have carried out a two loop computation of the low-energy effective action
for  the  four-dimensional $N=2$ supersymmetric Yang-Mills system coupled to
 hypermultiplets,  with  the chiral superfields of the
vector multiplet lying in an abelian subalgebra.  We have found a complete
cancellation at the level of the integrands of Feynman amplitudes, and
therefore the two loop  contribution to the action,
effective or Wilson, is identically zero.

\end{abstract}

\begin{flushright}
May 1997
\end{flushright}
\end{titlepage}

\newpage
Following the work by Seiberg and
Witten \cite{sw}  on the construction of the {\em exact} low-energy
effective action for the $N=2$ supersymmetric Yang-Mills $SU(2)$
 theory in its Coulomb phase, considerable attention has been given to
 the study of effective actions for $N=2$  theories. This
construction starts from a classical  action which is invariant under a global
$U(1)$ symmetry ($R$ invariance) and a local  $SU(2)$ symmetry. By giving
a nonvanishing vacuum expectation value to the chiral superfields of the
$N=2$ vector multiplet one both breaks the global invariance and 
induces a spontaneous symmetry breaking of $SU(2)$ to $ U(1)$.
The set of inequivalent vacua,  parametrized by the vacuum expectation value,
determines the "moduli space"; one  aims then  to obtain an effective
lagrangian description of the theory for any  value of the modulus.
While perturbative calculations are reliable in regions of the moduli
space corresponding to
weak coupling, in general quantum computations are difficult in the strong
coupling regime. The major breakthrough in the work by Seiberg and Witten has
been the realization that the complete non-perturbative answer could be
actually reconstructed if the perturbative contribution were known exactly.

In fact, $N=2$ supersymmetry fixes the general form of
the low energy effective action (i.e. that part of the effective action
which is leading for vanishing momenta): it is given in $N=2$ superspace
by a chiral integral of a holomorphic function $\FF(W)$, where $W$ is the
$N=2$ gauge superfield strength of the unbroken $U(1)$. In a $N=1$
superfield description it has the   general form \cite{ef}
 \beq
S_{eff}=\frac{1}{16\pi} {\cal I}m \left[\int d^4x \,d^2\th\, d^2{\bar\th}\,\,
\fb
\FF_{\f} (\f)
+ \int d^4x \,d^2\th\,\,
\FF_{\f \f}(\f )\, (\shalf W^\a W_\a)
\right]\ .
\label{S0}
\eeq
where $\f$ denotes the $N=1$ chiral superfield and $W_\a$ the $N=1$ gauge
superfield strength of the $N=2$, $U(1)$ gauge multiplet. One  then has to 
compute $\FF(\f)$. The work in ref. \cite{sw} is based in part on the fact that the 
perturbative contribution to the low energy effective action, i.e., to $\FF(\f)$,
is  one-loop exact  \cite{seiberg}. This simple result follows essentially from 
general arguments based on properties of the Wilson action \cite{sv}, and
the fact that the $N=2$ Yang-Mills theory has an exact one-loop beta function.
While this last feature is by now well established, the first point is more 
delicate and subtle. The Wilson action $S_W$ differs in general from the 
effective action $\G$ because in the
former one excludes infrared effects by construction, i.e., all loop-momenta
are integrated down to an infrared cutoff. In theories with massive particles only,
there is no important difference between $S_W$ and $\G$; however, 
when massless particles are present one has to keep   this distinction  in mind.

For the low energy effective theory of $N=2$ Yang Mills one would argue that
$S_W$ and $\G$ are equivalent
since the theory does not contain non-abelian gauge interactions and therefore
it is well behaved in the infrared. On the other hand it seems natural to ask
the question of what happens at the origin of the moduli
space, where the $SU(2)$ symmetry is restored and the non-abelian gauge bosons
become massless. Since the exact knowledge of the perturbative contribution
is crucial for the subsequent determination of the complete
nonperturbative result, and  to clarify the issue of effective action
versus Wilson action, we have studied the situation  at the two loop level.

In this Letter we describe an explicit, two-loop supergraph calculation
for the low energy effective action. We have found that  {\em complete} cancellation
occurs at the level of the integrands in the Feynman momentum integrals,
without the need  of any infrared or ultraviolet cutoffs to regularize
the theory. This unambiguously checks the complete equivalence of $S_W$ and
$\G$ for the case under consideration. Our calculation follows the
method used in ref. \cite{nonholo} where nonholomorphic contributions
to the effective action at the one--loop level were computed.
In contradistinction to that work, we  focus our attention on the low-energy
effective action of the massless fields by restricting the
 (external) chiral superfield component   of the $N=2$ vector multiplet to  
an abelian subalgebra, as in the Seiberg-Witten analysis.

The classical action for $N=2$ supersymmetric systems
written in $N=1$ superspace is
\bea{a}
S_{class} &=& \frac{1}{4g^2}\left[ \int d^4x \,d^2 \th\, {\rm tr}
(\shalf W^\a W_\a )+\int d^4x \,d^4 \th\,{\rm tr}( \Phibar \,e^{-V}
\Phi \, e^V ) \right] \nonumber \\
&&~~~~~~ +\int d^4x \,d^4 \th\, (\bar{Q}\, e^V Q +\tilde{Q} \, e^{-V}
\bar{\tilde Q}) + \Big(i\int d^4x\,d^2 \th\,\tilde Q \Phi Q +
h.c.\Big) \,,
\label{action}
\eea
The superfields $W_\a  \equiv i\bar D^2(e^{-V}D_\a e^V)$ and $\Phi$ are $N=1$
superfield components of the chiral $N=2$ superfield $W$ that describes
the $N=2$ vector multiplet.   $V$,  $\Phi$ and $\Phibar$ are Lie-algebra valued
in the adjoint representation, i.e. $V = V^a T_a$,
$\Phi= \Phi^a T_a$, etc.,  with $[T_a,T_b]= i f_{ab}^c T_c$ and ${\rm tr} T_a
T_b = K \d_{ab}$. The chiral $N=1$ superfields $Q$ and $\tilde Q$ together
describe $N=2$ hypermultiplets.  $Q$ and $\tilde Q$ are in mutually conjugate
representations $R$ and $\tilde R$. We shall explicitly consider the $SU(2)$ 
case, with $f_{abc} = \e_{abc}$ and $(T_a)_{kl} =i \e_{kal}$ for the adjoint 
representation and $(T_a)_{kl} =\frac{1}{2} ( \s_a)_{kl}$
for the fundamental representation.

In addition, after quantizing the vector multiplet  we   obtain the ghost
lagrangian
\beq
S_{ghost}  = \int d^4x \,d^4 \th~{\rm tr} \left[ \bar{c}'c - c' \bar{c}
+\frac{1}{2}(c'+\bar{c}')[V, c+ \bar{c}] + \cdots \right]
\label{sg}
\eeq
with the ghosts  in the adjoint representation. We use the notations and
conventions  of {\em Superspace} \cite{Superspace}.

We perform perturbative calculations by making a quantum-background splitting
$\Phi\rightarrow \Phi + {\phi}$, $\Phibar \rightarrow \Phibar +
\bar{\phi}$ and, for the $SU(2)$ case considered here, we choose the 
background   fields to lie  in the (abelian)  ``$z$'' direction,
\beq
{\phi} = (0,0,\phi ) ~~~~,~~~~\bar{\phi}= (0,0, \bar{\phi} )
\label{shift}
\eeq

We consider all diagrams with external
 $\phi$ and $\bar{\phi}$.  Contributions come from
internal lines corresponding to the $(Q,\tilde Q)$ hypermultiplets and
the $N=2$ vector multiplet itself.
Since we are interested in the low-energy effective action, when
doing $D$-algebra, we  drop terms with spinor
or space-time derivatives on the external lines. Also, as in ref.
\cite{nonholo} we work in Landau gauge for the $N=1$ vector multiplet.
This choice is particularly advantageous: it amounts to dropping
all cubic vertices of the form ${\rm tr} (\Phibar[V,\phi])$ and
${\rm tr}(\bar{\phi}[V,\Phi])$ since the $D^\a D^2 D_\a$ factor carried by the
$V$-propagator annihilates the $D^2$ or $\bar{D^2}$ factors from the
quantum $\Phibar$ or $\Phi$ line (and we are not considering terms where the
spinor derivatives are integrated by parts onto the background). Thus
for the quantum $V$ and $Q$ superfields  one can define
effective propagators obtained by summing over
arbitrary numbers of insertions of the external lines $\phi$.
We have then in the vector sector

\bea
{}<V^a V^b> &=& - \frac{4g^2}{K}\pmatrix{
\frac{1}{\raise-.04cm\hbox{$\scriptstyle p^2+\phi \bar{\phi}$}} & 0 & 0 \cr
0 & \frac{1}{\raise-.04cm\hbox{$\scriptstyle p^2+\phi \bar{\phi}$}} & 0 \cr
0 & 0 & \frac{1}{p^2} \cr
}^{ab}
\frac{D^\a \Dbar^2D_\a}{p^2} \d^{(4)}(\th - \th ') \nonumber\\
&& \nonumber\\
<\phibar^a \phi^b> &=& \frac{4g^2}{K} \d^{ab} \frac{1}{p^2} \d^{(4)}(\th -
\th ')\nonumber\\
&&\nonumber\\
<\bar{c}'^a c^b> &=&<\bar{c}^a c'^b> =
 \frac{4g^2}{K} \d^{ab} \frac{1}{p^2} \d^{(4)}(\th -
\th ')
\label{vprop}
\eea
For the hypermultiplet in  the adjoint representation we find
\bea
{}< \bar{Q}^a Q^b> &=& \frac{1}{K} \pmatrix{
\frac{1}{\raise-.04cm\hbox{$\scriptstyle p^2+\phi \bar{\phi}$}}  & 0 & 0 \cr
0 & \frac{1}{\raise-.04cm\hbox{$\scriptstyle p^2+\phi \bar{\phi}$}}  & 0 \cr
0 & 0 & \frac{1}{p^2} \cr
}^{ab} \d^{(4)}(\th - \th ')  \nonumber\\
&&\nonumber\\
<Q^a \tilde{Q}^b> &=& -\frac{1}{K} \pmatrix{
\frac{1}{\raise-.04cm\hbox{$\scriptstyle p^2+\phi \bar{\phi}$}} & 0 & 0 \cr
0 & \frac{1}{\raise-.04cm\hbox{$\scriptstyle p^2+\phi \bar{\phi}$}}  & 0 \cr
0 & 0 & \frac{1}{p^2} \cr
}^{ac}     \e^{c3b}~ \bar{\phi} ~\frac{D^2}{p^2}\d^{(4)}(\th - \th ')
\label{qaprop}
\eea
whereas in the fundamental representation
\bea
{}< \bar{Q}^i Q^j> &=& \frac{4\d^{ij}}{4p^2 +\phi \bar{\phi}} \d^{(4)}(\th - \th ')
\nonumber\\
&&\nonumber\\
<Q^i \tilde{Q}^j> &=& -i (\s^3)^{ij} \frac{2 \fb}{p^2(4p^2+\phi
\bar{\phi})}
D^2 \d^{(4)}(\th
- \th ')
\label{qfprop}
\eea
As usual, factors of $\bar{D}^2$ and $D^2$ appear  at  the ends of  chiral and
antichiral lines.

The one-loop contribution is by now a standard result. One obtains
a divergent contribution which simply leads to a renormalization of the
coupling constant $g$ in front of the action in (\ref{action}), as well 
as a holomorphic contribution to the function $\FF$ given by
\bea
{}\FF (\phi) = \frac{-i}{2\p} \left( {\rm tr}_R(\phi^2
\ln\frac{\phi^2}{\m^2})
- {\rm tr}_{ad} (\phi^2 \ln\frac{\phi^2}{\m^2}) \right)
\label{holo}
\eea
where $\m$ is the renormalization scale, and the two terms come
from the hypermultiplet and vector sectors respectively.
At one loop the ghosts do not contribute.

We turn now to the two-loop calculation. The general strategy is to draw
all possible supergraphs, evaluate first group theory factors in order to drop 
immediately the vanishing ones, then perform the $D$-algebra and assemble
the result in the form of momentum integrals. We list below separately the
contributions from diagrams which are not zero for group theory or
 $D$-algebra reasons.

We consider first the pure $N=2$ vector multiplet case. The relevant
supergraphs are given in Fig.1.
\vspace{0.2in}

\let\picnaturalsize=N
\def\picsize{5.0in}
\def\picfilename{2loop1.eps}
\ifx\nopictures Y\else{\ifx\epsfloaded Y\else\input epsf \fi
\let\epsfloaded=Y
\centerline{\ifx\picnaturalsize N\epsfxsize \picsize\fi
\epsfbox{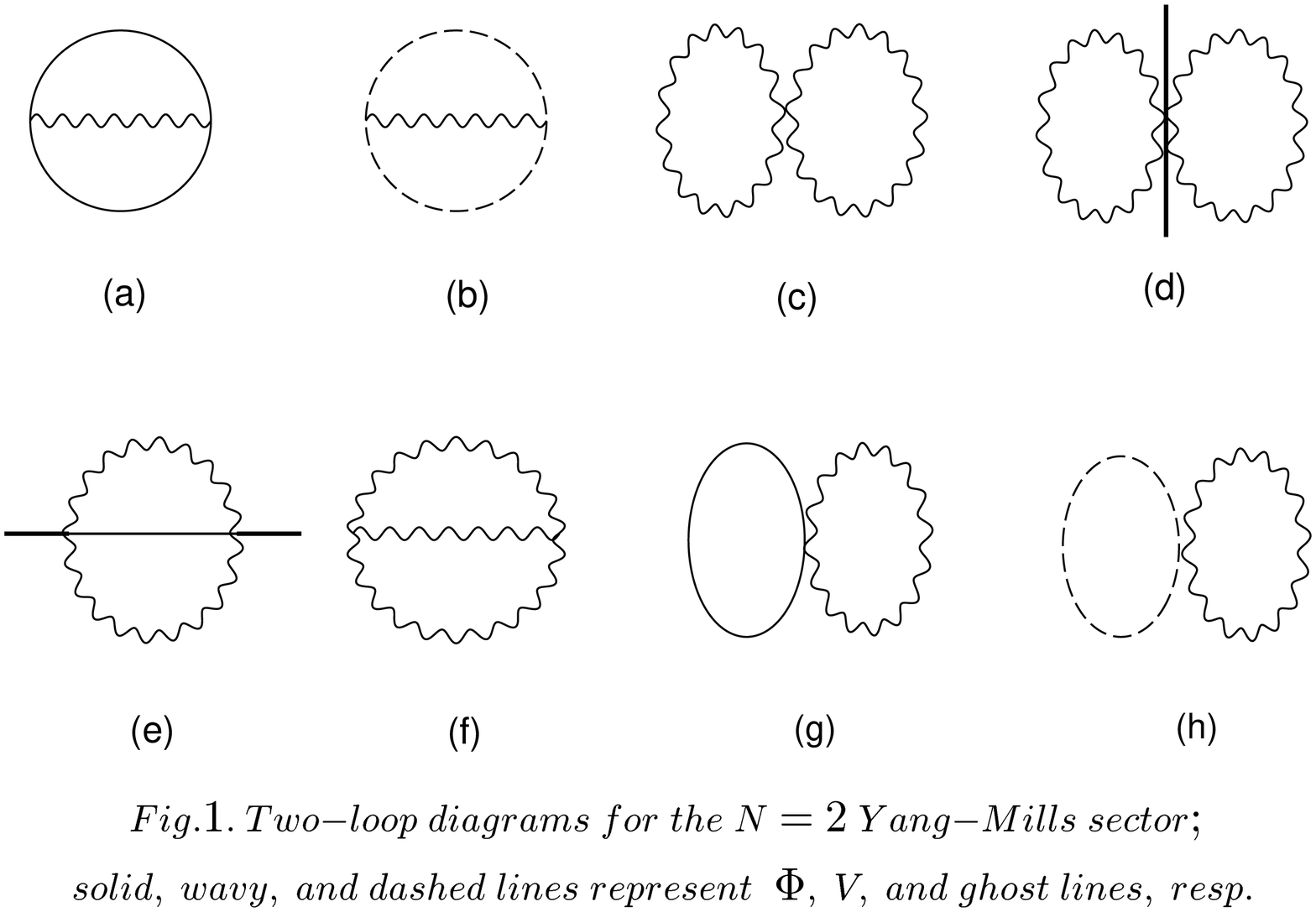}}}\fi
Note that in addition to the external $\phi$ lines which give rise to the
effective masses and are not indicated, Figs. 1d,e contain some  vertices
involving explicit  external  $\phi$ lines. After  $D$-algebra, we obtain the
following contributions, all multiplied by the common factor
$$
\frac{4g^2}{K} \int \frac{d^4k}{(2\pi)^4}  \frac{d^4p}{(2\pi)^4} d^4 \th
$$

\vspace{0.2in}
\bea
{}1a~~~ &:&~~~~-\frac{2 p \cdot k}{p^2 k^2 (p+k)^2} \left[
\frac{2}{(p+k)^2+\phi \bar{\phi}} +
\frac{1}{(p+k)^2} \right]  \nonumber\\
&&\nonumber\\
1b~~~&:&~~~~\frac{ p \cdot k}{p^2 k^2 (p+k)^2} \left[ \frac{2}{(p+k)^2+\phi
\bar{\phi}} +
\frac{1}{(p+k)^2} \right] \nonumber\\
&&\nonumber\\
1c~~~&:&~~~~ \frac{1}{3k^2}\left[ \frac{1}{(p^2+\phi
\bar{\phi})(k^2+\phi \bar{\phi})}
+\frac{1}{p^2(k^2+\phi \bar{\phi})} +\frac{1}{k^2(p^2+\phi
\bar{\phi})} \right] \nonumber\\
&&\nonumber\\
1d~~~&:&~~~~ \frac{\phi \bar{\phi}}{3k^2p^2}\left[ \frac{4}{(p^2+\phi
\bar{\phi})(k^2+\phi \bar{\phi})} +
\frac{1}{p^2(k^2+\phi \bar{\phi})} \right]  \nonumber\\
&&\nonumber\\
1e~~~&:&~~~~- \frac{  \phi \bar{\phi}~p \cdot k}{p^2k^2(p+k)^2} \left[
\frac{2}{(p^2+\phi \bar{\phi})(k^2+\phi \bar{\phi})}
+ \frac{1}{p^2(k^2+\phi \bar{\phi})} \right] \nonumber\\
&&\nonumber\\
1f~~~&:&~~~~ \frac{ p^2 k^2 - p \cdot k (p+k)^2}{2 p^2k^2 } \left[
\frac{1}{p^2(k^2+\phi \bar{\phi})[(p+k)^2+\phi \bar{\phi}]}
\right. \nonumber\\
&&\left.~~~~~~~~~~~~~~~~~~~+
\frac{1}{k^2(p^2+\phi \bar{\phi})[(p+k)^2+\phi \bar{\phi}]} +
\frac{1}{(p+k)^2(k^2+\phi \bar{\phi})(p^2+\phi \bar{\phi})} \right]
\nonumber\\
1g~~~&:&~~~~ -\frac{2}{k^2p^2} \left[ \frac{2}{k^2+\phi \bar{\phi}}
+\frac{1}{k^2}\right]
\nonumber\\
&&\nonumber\\
1h~~~&:&~~~~ \frac{2}{3}~\frac{1}{k^2p^2}
\left[ \frac{2}{k^2+\phi \bar{\phi}} +\frac{1}{k^2}\right]
  \label{ym}
\eea

If hypermultiplets are present, we get additional contributions
 presented in Fig. 2. In the result from Figs. 2a,b we have included similar
contributions from $\tilde{Q}$.

\vspace{0.4in}
\let\picnaturalsize=N
\def\picsize{4.5in}
\def\picfilename{2loop2.eps}
\ifx\nopictures Y\else{\ifx\epsfloaded Y\else\input epsf \fi
\let\epsfloaded=Y
\centerline{\ifx\picnaturalsize N\epsfxsize \picsize\fi
\epsfbox{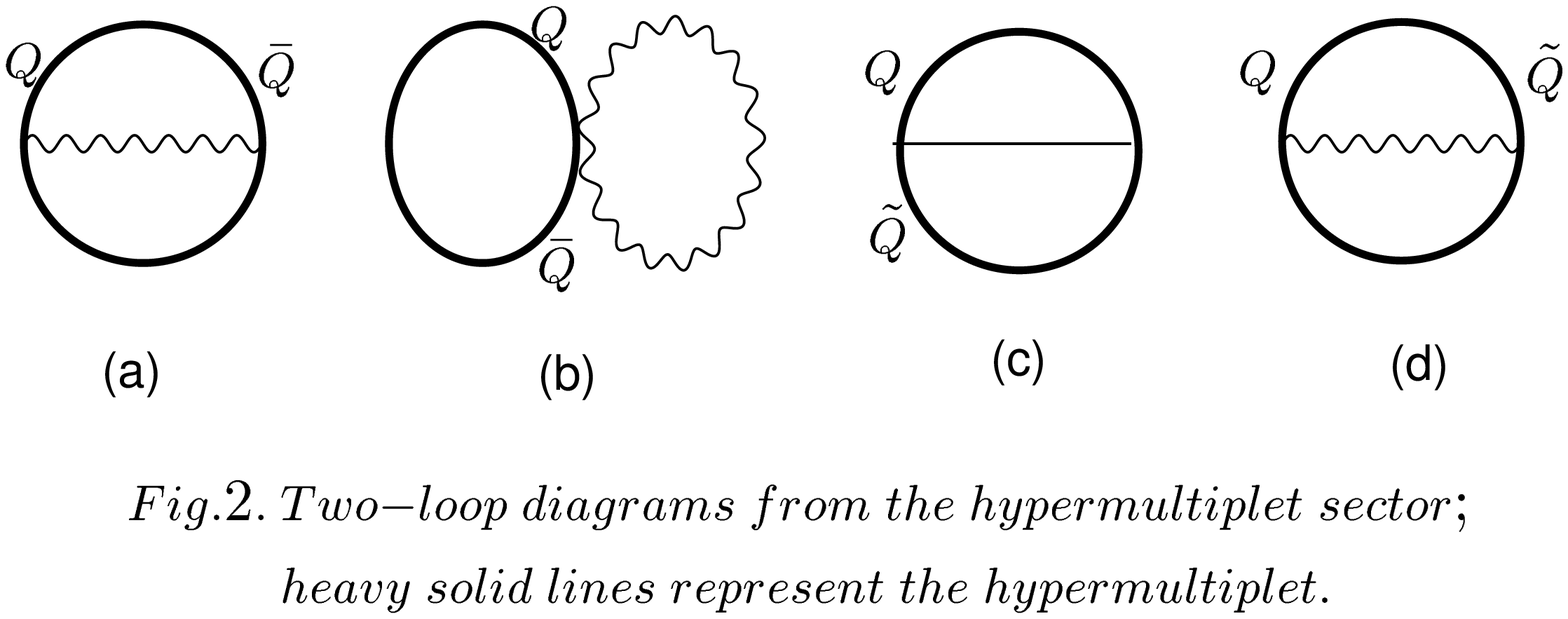}}}\fi

\vspace{0.2in}

If in the adjoint representation, we obtain
\bea  
{}2a~~~&:&~~~~-\frac{4 p \cdot k}{(p+k)^2} \left[ \frac{1}{(p^2+
\phi \bar{\phi})(k^2+\phi \bar{\phi})(p+k)^2}
+ \frac{2}{(p^2+\phi \bar{\phi})k^2[(p+k)^2+\phi \bar{\phi}]}
\right] \nonumber\\
&&\nonumber\\
2b~~~&:&~~~~-\frac{4}{k^2} \left[ \frac{1}{(p^2+\phi
\bar{\phi})(k^2+\phi \bar{\phi})}
+\frac{1}{p^2(k^2+\phi \bar{\phi})} +\frac{1}{k^2(p^2+\phi
\bar{\phi})}  \right] \nonumber\\
&&\nonumber\\
2c~~~&:&~~~~\frac{2}{(p+k)^2}  \left[ \frac{1}{(p^2+\phi
\bar{\phi})(k^2+\phi \bar{\phi})}
+\frac{1}{p^2(k^2+\phi \bar{\phi})} +\frac{1}{k^2(p^2+\phi
\bar{\phi})}  \right] \nonumber\\
&&\nonumber\\
2d~~~&:&~~~~ \frac{4\phi \bar{\phi}} {(p^2+\phi
\bar{\phi})(k^2+\phi \bar{\phi})(p+k)^4}
\label{adj}
\eea

If in the fundamental representation,
\bea
{}2a~~~&:&~~~~-\frac{16 p \cdot k}{(4p^2+\phi \bar{\phi})(4k^2+\phi
\bar{\phi})(p+k)^2} \left[
\frac{2}{(p+k)^2+\phi \bar{\phi}} + \frac{1}{(p+k)^2} \right] \nonumber\\
&&\nonumber\\
2b~~~&:&~~~~-\frac{4}{(4p^2+\phi \bar{\phi})k^2} \left[
\frac{2}{k^2+\phi \bar{\phi}} + \frac{1}{k^2}
\right] \nonumber\\
&&\nonumber\\
2c~~~&:&~~~~ \frac{24}{(4p^2+\phi \bar{\phi})(4k^2+\phi
\bar{\phi})(p+k)^2} \nonumber\\
&&\nonumber\\
2d~~~&:&~~~~\frac{4\phi \bar{\phi}}{(4p^2+\phi
\bar{\phi})(4k^2+\phi \bar{\phi})(p+k)^2} \left[\frac{1}{(p+k)^2} -
\frac{2}{(p+k)^2+\phi \bar{\phi} }\right] \label{fund}
\eea

It is then a matter of straightforward algebra to check that separately, the
contributions in (\ref{ym}), (\ref{adj}) and (\ref{fund}) add up to zero.
We have thus verified by an explicit supergraph calculation that the two-loop
contribution to the low-energy  $N=2$ effective action for the pure Yang-Mills
theory, or for the Yang-Mills theory coupled to hypermultiplets, vanishes 
identically.

Some comments are in order:

\noindent a. The cancellation of the  contributions to the low energy
effective action is complete, including tadpole-type  $1/p^2$ integrals that we
would normally discard in dimensional regularization; there is no need for
 some low-energy cutoff.  In principle,   we should imagine some ultraviolet
regularization procedure (such as dimensional regularization) which permitted
us to shift loop momenta as needed, in order to obtain the cancellation. 
Presumably, even this is not necessary.

\noindent b. The removal of divergences of the $N=2$ effective action amounts
to a one-loop renormalization (in dimensional regularization, say)
 $ 1/g^2 \rightarrow 1/g^2 + c / \e$. Consequently, the coupling constant $g$
itself  gets renormalized at all orders,
 $g^2 \rightarrow g^2 - c g^4/\e +c^2 g^6/\e^2 + \cdots...$,
as expected from the 't Hooft pole equations.

\noindent c. At the one-loop level, in ref. \cite{nonholo} we have found some
non-holomorphic corrections to the effective action when the external chiral
 superfields $\phi$ do not
lie in an abelian subalgebra. It is of interest to examine the corresponding
situation at the two-loop level.

\begin{flushleft}
{\bf Acknowledgement}
\end{flushleft}

We are happy to thank numerous colleagues for discussions.
 MG and MR acknowledge NSF Grants Nos.\ PHY-92-22318 and 93-09888 
for partial support.

\end{document}